\documentclass[twocolumn,prc,showpacs,showkeys]{revtex4}
\usepackage{amsfonts}
\usepackage{amsmath}
\usepackage{amssymb}
\usepackage{graphicx}
\usepackage{rotating}

\providecommand{\U}[1]{\protect\rule{.1in}{.1in}}
\providecommand{\U}[1]{\protect\rule{.1in}{.1in}}
\providecommand{\U}[1]{\protect\rule{.1in}{.1in}}
\providecommand{\U}[1]{\protect\rule{.1in}{.1in}}
\providecommand{\U}[1]{\protect\rule{.1in}{.1in}}

\begin{document}

\title{Resonance-like nuclear processes in solids: 3rd and 4th order
processes\\
}
\author{P\'{e}ter K\'{a}lm\'{a}n\footnote{%
retired from Budapest University of Technology and Economics, Institute of
Physics \newline
e-mail: kalmanpeter3@gmail.com}}
\author{Tam\'{a}s Keszthelyi}
\affiliation{Budapest University of Technology and Economics, Institute of Physics,
Budafoki \'{u}t 8. F., H-1521 Budapest, Hungary\ }

\begin{abstract}
It is recognized that in the family of heavy charged particle and electron
assisted double nuclear processes resonance-like behavior can appear. The
transition rates of the heavy particle assisted 3rd-order and electron
assisted 4th-order resonance like double nuclear processes are determined.
The power of low energy nuclear reactions in $Ni-H$ systems formed in $Ni$
placed in $H_{2}$ gas environment is treated. Nuclear power produced by
quasi-resonant electron assisted double nuclear processes in these $Ni-H$
systems is calculated. The power obtained tallies with experiments and its
magnitude is considerable for practical applications.
\end{abstract}

\volumenumber{number}
\issuenumber{number}
\eid{identifier}
\date[Date text]{date}
\received[Received text]{date}
\revised[Revised text]{date}
\accepted[Accepted text]{date}
\published[Published text]{date}
\startpage{1}
\endpage{}
\maketitle

\section{Introduction}

Since the "cold fusion" publication by Fleischmann and Pons in 1989 \cite%
{FP1} a new field of experimental physics has emerged. Although even the
possibility of the phenomenon of nuclear fusion at low energies is doubted
by many representatives of mainstream physics, the quest for low-energy
nuclear reactions (LENR) flourished and hundreds of publications (mostly
experimental) have been devoted to various aspects of the problem. (For the
summary of experimental observations, the theoretical efforts, and
background events see e.g. \cite{Krivit}, \cite{Storms2}.) The main reasons
for revulsion against the topic have been: (a) according to standard
knowledge of nuclear physics due to the Coulomb repulsion no nuclear
reaction should take place at energies corresponding to room temperature,
(b) the observed extra heat attributed to nuclear reactions is not
accompanied by the nuclear end products expected from hot fusion
experiences, (c) nuclear transmutations were also observed, that considering
the Coulomb interaction is an even more inexplicable fact at these energies.

The situation is further complicated by the fact that the electrolysis, gas
discharge and/or high pressure gas environments that are stipulated to
induce LENR have their effect through rather complex microscopic processes
that in most of the cases are difficult to reproduce. As a result, for
explaining the riddle of cold fusion it is indispensable to understand
theoretically the underlying nuclear reactions.

In our opinion we made progress \cite{KaKe} in the theoretical clarification
of the nuclear physics behind LENR. The idea is based on the fact that if a
heavy, charged particle (proton, deuteron) of low energy enters a solid
(metal), then through the Coulomb interaction it changes the state of
charged particles, primarily quasi free valence electrons in the metal while
its own state is also changed. Our results of standard perturbation theory
calculations indicate that by means of the Coulomb interaction the ingoing
charged particle in this change of state can obtain so high value of virtual
momentum (energy) that is enough to induce various nuclear reactions
including fusions and/or transmutations. With the results of our theoretical
work, we found reasonable answers to the above questions (a, b and c).

In \cite{KaKe} however, it was not taken into account that the emergence of
new charged particles may alter the state of the solid too in a way that can
allow further nuclear processes, such as the family of double nuclear
processes which has special interest from the point of view of low energy
nuclear processes. In this paper this effect is considered. First a
3rd-order double nuclear process is investigated the graphs of which can be
seen in FIG. 1. It is shown that a resonance can appear in this process.
Next a 4th-order double nuclear process is discussed the graphs of which can
be seen in FIG. 2 which shows resonance-like characters too. Finally the
nuclear power produced in $Ni-H$ systems is calculated.

This paper is mainly based on \cite{KaKe}. The notation and the outline of
the calculation is the following. The Coulomb coupling strength $%
e^{2}=\alpha _{f}\hbar c$ and the strong coupling strength $f^{2}=0.08\hbar
c $ \cite{Bjorken}, $e$ is the elementary charge, $\alpha _{f}$ is the fine
structure constant, $\hbar $ is the reduced Planck constant and $c$ is the
velocity of light.

When calculating the matrix elements of the strong interaction potential,
the long wavelength approximation $\left\vert \varphi (\mathbf{0}%
)\right\vert =f_{jk}(k)/\sqrt{V}$ of the Coulomb solution $\varphi (\mathbf{r%
})$ is used, that is valid in the range of a nucleon, where $f_{jk}(k)$ is
the appropriate Coulomb factor corresponding to the particles, which take
part in strong interaction and $V$ is the volume of normalization. We
introduced the following notation 
\begin{equation}
F_{jk}(k)=f_{jk}^{2\text{ }}(k)=\frac{2\pi \eta _{jk}\left( k\right) }{\exp %
\left[ 2\pi \eta _{jk}\left( k\right) \right] -1}.  \label{DCbh}
\end{equation}%
Here the Sommerfeld parameter $\eta _{jk}\left( k\right) $ for particles $j$
and $k$ of electric charge numbers $z_{j}$ and $z_{k}$ is determined as 
\begin{equation}
\eta _{jk}\left( k\right) =z_{j}z_{k}\alpha _{f}\frac{\mu _{jk}c}{\hbar k},
\label{eta23}
\end{equation}%
where $k=\left\vert \mathbf{k}_{j}-\mathbf{k}_{k}\right\vert $ is the
magnitude of the relative wave vector $\mathbf{k=k}_{j}-\mathbf{k}_{k}$ of
the interacting particles of wave vectors $\mathbf{k}_{j}$ and $\mathbf{k}%
_{k}$ and $\mu _{jk}=m_{j}m_{k}/\left( m_{j}+m_{k}\right) $ is the reduced
mass of particles of rest masses $m_{j}$ and $m_{k}$.

For quasi-free particles (electron and ingoing proton) taking part in
Coulomb interaction we use plane waves. Thus their Coulomb matrix element is
calculated in the Born approximation which is corrected with the so called
Sommerfeld factor 
\begin{equation}
g_{S}\left( k_{in},k_{out}\right) =f_{jk}(k_{in})/f_{jk}(k_{out})
\end{equation}%
\cite{Heitler}, where $k_{in}$ and $k_{out}$ are the magnitudes of the
relative wave numbers before and after Coulomb scattering. For other details
and notation see \cite{KaKe}.

\section{Resonance-like heavy charged particle and electron assisted double
nuclear processes}

Preliminarily one must emphasize that in both (3rd and 4th order) double
nuclear processes discussed resonances may occur. The reason for the
possibility of resonance is that the continuum of the kinetic energy of an
intermediate state is shifted down by the energy of the (first) nuclear
transition and therefore one of the denominators in the perturbation
calculation can be equal to zero. The occurrence of resonances increases the
rate significantly.

\subsection{Resonance-like heavy charged particle assisted double nuclear
processes}

First the resonance-like heavy charged particle assisted double nuclear
processes (see FIG. 1) are discussed. The particles in FIG. 1 are all heavy,
and positively charged. The ingoing particle is particle 2, which belongs to
system $B$ (the ensemble of incoming particles forms system $B$). It is
supposed that it has moderately low energy (of about $keV$ order of
magnitude), that is raised in the second order processes discussed in \cite%
{KaKe} or after it in the decelerating process. Particle 2 scatters by
Coulomb scattering on particle 1 localized in the solid (system $A$).
Particles 3 and 4 are the nuclear targets of system $A$. The particles in
the intermediate state 1' and 2' may pick up large enough wave vector to
overcome the Coulomb repulsion due to particles 4 and 3. Particles 5 and 6
are products of the process. The nuclear process $1^{\prime }+4\rightarrow 5$
takes place as the consequence of the modification of system $A$ by system $%
B $. Here both nuclear processes are thought to be nuclear captures. It can
be shown (see Appendix I.) that the process may have resonance like
character if the masses of particles 5 and 6 differ significantly, therefore
the contribution of the leading graph (e.g. FIG. 1(a) in the case $m_{5}\gg
m_{6} $ discussed) to the rate is enough to calculate.

\begin{figure}[tbp]
\resizebox{6.0cm}{!}{\includegraphics*{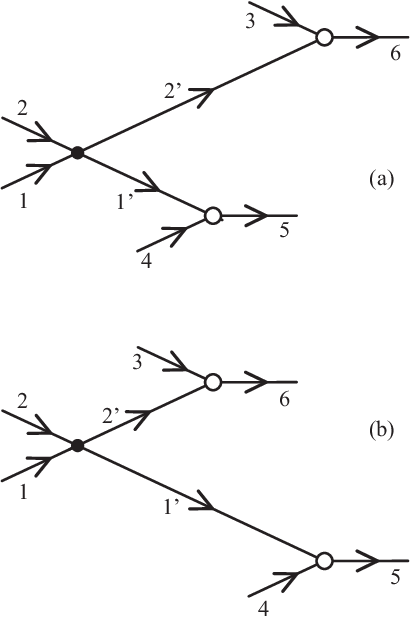}}
\caption{The graphs of quasi-resonant heavy charged particle
Coulomb-assisted nuclear reactions. The filled dot denotes
Coulomb-interaction and the open circle denotes nuclear (strong)
interaction. Free particle 2 (e.g. p or d) belongs to system B. Localized
particles 1, 3 and 4 belong to system A. All the particles are heavy and
positively charged. From the point of view of the nuclear process particles
1' and 2' are ingoing, particles 3 and 4 are targets and particles 5 and 6
are reaction products. In the case discussed in the text in the process of
FIG. 1(a) there is a resonance, therefore its contribution to the rate is
dominant and the contribution of the process of FIG. 1(b) is neglected.}
\label{figure1}
\end{figure}

We take as initial state of a localized particle $k$ 
\begin{equation}
\psi _{ki}\left( \mathbf{x}\right) =\left( \beta _{k}^{2}/\pi \right)
^{3/4}\exp \left( -\beta _{k}^{2}\mathbf{x}^{2}/2\right) ,  \label{psziki}
\end{equation}%
which is the ground state of a 3-dimensional harmonic oscillator of energy $%
E_{0}=\frac{3}{2}\hbar \omega _{k}$ and of angular frequency $\omega _{k}$ 
\cite{Cohen}. Now we take particle 1 $\left( k=1\right) $ localized and the
initial state $\psi _{1i}\left( \mathbf{x}\right) $ is of the form $\left( %
\ref{psziki}\right) $ with the parameter $\beta _{1}=\sqrt{m_{1}\omega
_{1}/\hbar }$, where $m_{1}$ is the rest mass of the localized particle 1.
The Coulomb matrix elements which contain initial state of form $\psi
_{ki}\left( \mathbf{x}\right) $ also preserve momentum \cite{KaKe}.

The total rate of the reaction

\begin{eqnarray}
W_{tot}^{(3)} &=&K^{(3)}K_{1}^{(3)}\left( \Delta ,\Delta _{5}\right)
h_{corr,3}^{2}h_{corr,4}^{2}N_{134}  \label{Wtot3} \\
&&\times \sum_{k_{2i}}H\left( k_{2i},k^{r}\right) N_{2}\left( k_{2i}\right) ,
\notag
\end{eqnarray}%
where $N_{134}$ is the number of triples of particles 1, 3 and 4, that may
be effective for one particle 2, and $N_{2}\left( k_{2i}\right) $ is the
actual number of quasi-free particles 2 of wave number $k_{2i}$. Furthermore 
\begin{equation}
K^{(3)}=144\alpha _{f}^{2}\pi ^{-5/2}\left( 1-\frac{2}{e}\right) ^{4}\left( 
\frac{f^{2}}{\hbar c}\right) ^{4}R^{2}c\beta _{1}^{3},  \label{K3}
\end{equation}%
where $R=1.2\times 10^{-13}\left[ cm\right] $ is the proton-radius,%
\begin{equation}
K_{1}^{(3)}\left( \Delta ,\Delta _{5}\right) =\frac{\mu _{12}^{2}c^{4}\sqrt{%
\mu _{56}c^{2}\Delta }}{\sqrt{2}\left( \Delta _{5}-\frac{\mu _{56}}{m_{5}}%
\Delta \right) ^{3}}\frac{\mu _{56}}{m_{2}},  \label{K13}
\end{equation}%
where $\Delta _{5}=\Delta _{01}+\Delta _{04}-\Delta _{05}$ is the energy of
reaction $1^{\prime }+4$ $\rightarrow 5$, and $\Delta _{6}=\Delta
_{02}+\Delta _{03}-\Delta _{06}$ is the energy of reaction $2^{\prime
}+3\rightarrow 6$. The $\Delta _{0j}$-s are the energy defects of the
corresponding nuclei and $\Delta =\Delta _{5}+\Delta _{6}$ is the total
reaction energy. 
\begin{equation}
H\left( k_{2i},k^{r}\right) =g_{S}^{2}\left( k_{2i},2k^{r}\right)
F_{23}(k^{r})F_{14}(k^{r}),  \label{Hkr}
\end{equation}%
where 
\begin{equation}
k^{r}=\frac{\sqrt{2m_{2}\left( \Delta _{5}-\frac{\mu _{56}}{m_{5}}\Delta
\right) }}{\hbar }.  \label{kres}
\end{equation}%
$h_{corr,3}$, $h_{corr,4}$ are defined by Eq.(45) of \cite{KaKe} and are
determined as%
\begin{equation}
h_{corr,k}=A_{k}-z_{k}  \label{hcorrk}
\end{equation}%
in the case of proton captures, and 
\begin{equation}
h_{corr,k}=A_{k}  \label{hcorrk2}
\end{equation}%
in the case of deuteron captures. (For the details of the calculation see
Appendix. I.)

In a numerical example let particles 1, 2, 3 be deuterons, particle 4 be $%
_{46}^{106}Pd$, particle 5 be $_{47}^{108}Ag$ and particle 6 be $_{2}^{4}He$%
. If particle 1 is a deuteron then $\beta _{1}=4.81\times 10^{8}$ $\left[
cm^{-1}\right] $ (see Sec. VIII. A of \cite{KaKe}). With these choices 
\begin{equation}
\Delta _{6}=23.847\left[ MeV\right] ,\Delta _{5}=10.833\left[ MeV\right] ,
\end{equation}%
\begin{equation}
\Delta =34.681\left[ MeV\right] ,
\end{equation}%
\begin{equation}
\mu _{12}c^{2}=m_{0}c^{2}=931.494\left[ MeV\right] ,\text{ }\mu
_{56}c^{2}\simeq 4m_{0}c^{2},
\end{equation}%
\begin{equation}
\mu _{56}/m_{5}=1/28,\text{ }\mu _{56}/m_{2}\simeq 2.
\end{equation}%
With these numbers one can obtain 
\begin{equation}
\Delta _{5}-\frac{\mu _{56}}{m_{5}}\Delta =9.60\left[ MeV\right] ,
\end{equation}%
\begin{equation}
K_{1}^{(3)}\left( \Delta ,\Delta _{5}\right) =5.00\times 10^{5}
\end{equation}%
and 
\begin{equation}
\hbar ck^{r}=189\left[ MeV\right] ,
\end{equation}%
the latter producing%
\begin{equation}
F_{23}(k^{r})=0.89,F_{14}(k^{r})=2.79\times 10^{-8}.
\end{equation}%
Finally one gets 
\begin{equation}
W_{tot}^{(3)}=2360\times N_{134}\sum_{k_{2i}}g_{S}^{2}\left(
k_{2i},2k^{r}\right) N_{2}\left( k_{2i}\right) \left[ s^{-1}\right] .
\end{equation}%
The $g_{S}^{2}\left( k_{2i},2k^{r}\right) $ dependence is similar to the $%
G_{S}\left( k_{2i},2k_{1f}\right) $ dependence discussed at the end of
section VIII.B of \cite{KaKe}. For the number $N_{134}$, in the cases
discussed, as a lowest estimation 
\begin{equation}
N_{134}=\left( V_{eff}/v_{c}\right) r_{A_{1}}r_{A_{3}}r_{A_{4}},
\label{N134}
\end{equation}%
where $r_{A_{i}}$ is the relative natural abundance of isotopes of mass
number $A_{i}$ or $r_{A_{i}}=u$ if the $i$-th particle is a hydrogen
isotope, $V_{eff}$ is the volume effectively felt by a particle 2, $u$ is
the proton (or deuteron) over metal number density and $v_{c}\left(
=d^{3}/4\right) $ is the unit cell of the solid (in the case of $fcc$
metals).

It should be noted that there may be a great variety of other types of
possible quasi-resonant heavy charged particle assisted double nuclear
reactions the discussion of which is not given here.

\subsection{Resonance-like electron assisted double nuclear processes}

From the processes discussed up till now one can conclude the following: (a)
it is advantageous, if the wave vector (momentum) transferred through the
intermediate state by Coulomb interaction has the possible maximum value,
(b) the electron assisted process is advantageous since the Coulomb and
Sommerfeld factors cause minimal or negligible hindering in this case and
(c) the appearance of resonance significantly increases the rate. These
conclusions led us to a 4th-order, electron assisted doubled nuclear
process, the graphs of which can be seen in FIG. 2. At a particular choice
of the participants (see Appendix II.) resonances can be found in the
processes of FIG. 2 (a) and (b). The resonance arises in line 4'. Moreover,
the details of the calculation show that high contribution to the rate is
obtained if the energy of the electron is negligible in the final state
compared with the energy of the total reaction. In other words, the main
contribution to the rate is produced by final states in which particles 6
and 7 share the reaction energy.

\begin{figure}[ptb]
\resizebox{6.0cm}{!}{\includegraphics*{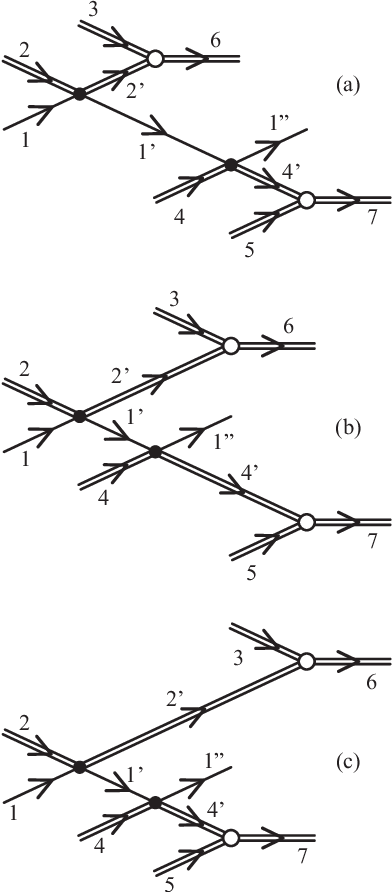}}
\caption{The graphs of quasi-resonant, electron assisted nuclear processes.
The simple lines represent free (initial (1), intermediate (1') and final
(1")) electrons. The doubled lines represent heavy, charged initial (free 2,
bound 4) particles (such as p, d), their intermediate states (2',4'), target
nuclei (3,5) and reaction products (6,7). The filled dot denotes
Coulomb-interaction and the open circle denotes nuclear (strong)
interaction. }
\label{figure2}
\end{figure}

It is supposed that particle 1 (1' and 1\textquotedblright ) is a quasi-free
electron of the solid (a metal). Particle 2 is an ingoing free, heavy,
positively charged particle of system $B$. Particles 3, 4, and 5 are heavy
particles of positive charge that are localized in the solid. If processes
of FIG. 2 (a) and (b) have resonance-like character (see Appendix II.) then
the process of FIG. 2 (c) has not and therefore its contribution may be
neglected. Now particle 4 is localized and its state is $\psi _{4i}\left( 
\mathbf{x}\right) $ given by $\left( \ref{psziki}\right) $.

The total rate $W_{tot}^{(4)}$ (for the details see Appendix II.) can be
obtained from the rate $W_{fi}^{(4)}$(see $\left( \ref{Wfi42}\right) $) as 
\begin{equation}
W_{tot}^{(4)}=K_{0}^{(4)}\left\langle F_{G}\right\rangle \chi \left( \Delta
\right) h_{corr,3}^{2}h_{corr,5}^{2}N_{1}N_{2}\frac{N_{345}}{V},
\label{Wtot4}
\end{equation}%
where $N_{1}$ is the instantaneous number of the quasi-free electrons that
are felt by one ingoing particle 2, $N_{2}$ is the number of quais-free
ingoing particles 2, $N_{345}/V$ is the number density of the target triples
of particles 3, 4 and 5, 
\begin{eqnarray}
K_{0}^{(4)} &=&2^{12}3\pi ^{11/2}\alpha _{f}^{4}\left( \frac{f^{2}}{\hbar c}%
\right) ^{4}\left( 1-\frac{2}{e}\right) ^{4}\left(
z_{1}^{2}z_{2}z_{4}\right) ^{2}  \label{K04} \\
&&\times \left( \hbar c\right) ^{6}R^{2}ck_{1,Max}^{3}\beta _{4}^{3},  \notag
\end{eqnarray}%
where $k_{1,Max}$ is the maximum of the possible wave vectors of the
electron in the final state. (We calculate the rate of those processes in
which the kinetic energy $E_{1f}$ of the electron can be neglected in the
energy $E_{f}$ of the final state). 
\begin{equation}
\left\langle F_{G}\right\rangle =\frac{\left\langle
F_{12}(k_{i})\right\rangle }{F_{12}(2k_{4}^{r})}F_{23}\left(
k_{4}^{r}\right) F_{45}\left( k_{4}^{r}\right) ,  \label{FG}
\end{equation}%
with%
\begin{equation}
k_{4}^{r}=\frac{\sqrt{2m_{4}\delta _{a,b}}}{\hbar }=\frac{\sqrt{2m_{4}\left(
\Delta _{6}-\frac{m_{7}}{m_{6}+m_{7}}\Delta \right) }}{\hbar }  \label{k4res}
\end{equation}%
Now $\Delta _{6}=\Delta _{02}+\Delta _{03}-\Delta _{06}$ is the energy of
reaction $2^{\prime }+3\rightarrow 6$ and $\Delta _{7}=\Delta _{05}+\Delta
_{04}-\Delta _{07}$ is the energy of reaction $4^{\prime }+5$ $\rightarrow 7$%
. The $\Delta _{0j}$-s are again the energy defects of the corresponding
nuclei and the total reaction energy%
\begin{equation}
\Delta =\Delta _{6}+\Delta _{7}.  \label{deltat4}
\end{equation}%
\begin{equation}
\chi \left( \Delta \right) =\frac{m_{7}}{m_{4}}\frac{\sqrt{2m_{7}c^{2}\Delta 
}}{\delta _{a,b}^{3}\Delta _{ab}^{2}}\left( \frac{1}{\Delta _{a}}+\frac{1}{%
\Delta _{b}}\right) ^{2},  \label{khi}
\end{equation}%
where $\delta _{a,b}$ is given by $\left( \ref{deltab}\right) $,
furthermore, for $\Delta _{a}$, $\Delta _{b}$ and $\Delta _{ab}$ and see $%
\left( \ref{Enia4d}\right) $, $\left( \ref{Enuib4d}\right) $ and $\left( \ref%
{deltab2}\right) $, and for $h_{corr,3}$ and $h_{corr,5}$ see $\left( \ref%
{hcorrk}\right) $ and $\left( \ref{hcorrk2}\right) $.

It is reasonable to take 
\begin{equation}
\frac{N_{345}}{V}=\frac{1}{v_{c}}r_{A_{3}}r_{A_{4}}r_{A_{5}}=\frac{4}{d^{3}}%
r_{A_{3}}r_{A_{4}}r_{A_{5}},  \label{N345}
\end{equation}%
where $v_{c}=d^{3}/4$ is the unit cell of the solid in the case of $fcc$
metals, $r_{A_{i}}$ is the relative natural abundance of isotopes of mass
number $A_{i}$ or $r_{A_{i}}=u$ if the $i$-th particle is a hydrogen
isotope. Thus 
\begin{equation}
W_{tot}^{(4)}=(\frac{4}{d^{3}}K_{0}^{\left( 4\right) })\left\langle
F_{G}\right\rangle \chi \left( \Delta \right)
h_{corr,3}^{2}h_{corr,5}^{2}N_{1}N_{2}r_{A_{3}}r_{A_{4}}r_{A_{5}}.
\label{Wtot42}
\end{equation}

If particle 4 is a deuteron then $\beta _{4}=4.81\times 10^{8}$ $\left[
cm^{-1}\right] $. Taking $E_{1,Max}=0.1$ $\left[ MeV\right] $, that is the
maximum of the possible energies of the electron in the final state, one
gets $k_{1,Max}^{3}=4.89\times 10^{30}$ $\left[ cm^{-3}\right] $ and 
\begin{equation}
\frac{4}{d^{3}}K_{0}^{\left( 4\right) }=5.7\times 10^{-8}\left(
z_{1}^{2}z_{2}z_{4}\right) ^{2}\text{ }\left[ MeV^{6}s^{-1}\right] 
\label{K14 d4}
\end{equation}%
in the case of $Pd$, i.e. with $d=3.89\times 10^{-8}$ $\left[ cm\right] $.

As a numerical example let particles 2, 4 and 5 be deuterons, particle 3 a $%
Pd$ isotope (of $A_{3}=106$ with $r_{A_{3}}=0.2733$) and particle 7 $%
_{2}^{4}He$. With this choice 
\begin{equation}
\Delta _{6}=10.833\left[ MeV\right] ,\Delta _{7}=23.847\left[ MeV\right] ,
\end{equation}%
\begin{equation}
\Delta =34.681\left[ MeV\right] ,
\end{equation}%
furthermore 
\begin{equation}
\delta _{a,b}=9.60\left[ MeV\right] ,\Delta _{ab}=198.7\left[ MeV\right] ,
\end{equation}%
\begin{equation}
\Delta _{a}=19.2\left[ MeV\right] ,\Delta _{b}=179.5\left[ MeV\right] ,
\end{equation}%
and $m_{7}/m_{4}=2$ resulting%
\begin{equation}
\chi \left( \Delta \right) =1.69\times 10^{-6}\left[ MeV^{-6}\right] .
\end{equation}%
\begin{equation}
F_{23}\left( k_{4}^{r}\right) =2.79\times 10^{-8},F_{45}\left(
k_{4}^{r}\right) =0.89,F_{12}(2k_{4}^{r})=1
\end{equation}%
and%
\begin{equation}
\left\langle F_{12}(k_{i}(E_{i}))\right\rangle =23.18\times \left\langle
E_{i}^{-1/2}\left[ eV\right] \right\rangle _{av}
\end{equation}%
producing 
\begin{equation}
\left\langle F_{G}\right\rangle =5.76\times 10^{-7}\times \left\langle
E_{i}^{-1/2}\left[ eV\right] \right\rangle _{av},
\end{equation}%
where $E_{i}$ is the energy of the initial conduction electron. Averaging $%
E_{i}^{-1/2}$ by means of the Fermi-Dirac distribution in the Sommerfeld
free electron model at $T=0$ yields 
\begin{equation}
\left\langle E_{i}^{-1/2}\right\rangle _{av}=2\left( E_{F}^{-1/2}\left[ eV%
\right] \right) 
\end{equation}%
where $E_{F}$ denotes the Fermi energy \cite{Solyom}. With these numbers 
\begin{equation}
W_{tot}^{(4)}=1.4\times 10^{-15}\left( E_{F}^{-1/2}\left[ eV\right] \right)
\times N_{1}N_{2}u^{2}\left[ s^{-1}\right] .
\end{equation}%
This rate produces a total power $P_{tot}^{(4)}=C_{E}W_{tot}^{(4)}\Delta $,
where $C_{E}=1.602\times 10^{-13}$ $\left[ J/MeV\right] $ is the energy unit
conversion factor. With a deuteron concentration independent $E_{F}=17\left[
eV\right] $, 
\begin{equation}
P_{tot}^{(4)}=1.8\times 10^{-27}\times N_{1}N_{2}u^{2}\left[ W\right] .
\label{Ptot4}
\end{equation}

In all the charged particle assisted processes discussed the quasi-resonant,
4th-order electron assisted double nuclear process seems to be the leading
one.

\section{Nuclear power in Ni-H systems due to double proton capture}

In this section we deal with a family of 4th-order resonant,
electron-assisted double nuclear processes shown in FIG. 2. Our aim is to
show that in $Ni-H$ systems formed in hydrogen gas one of the family of
processes in FIG. 2. may have high rate and the power generated by these
nuclear processes is also considerable from practical point of view that can
be calculated in our theory. In $Ni-H$ systems formed in hydrogen gas extra
heat production was observed \cite{Focardi1}, \cite{Focardi2} whose nuclear
origin was proven with neutron detection \cite{Battaglia}. Since in $Ni-H$
systems formed in $H$ gas there is no light particle for significant nuclear
effect save the natural deuteron content of hydrogen, according to our
theory the primary process to generate considerable energy is the
quasi-resonant electron assisted (double) proton capture of the $Ni$
isotopes.

The two proton captures%
\begin{equation}
\text{ }_{28}^{A}Ni+p\rightarrow \text{ }_{29}^{A+1}Cu+\Delta ,  \label{NiAp}
\end{equation}%
which are coupled due to the quasi-resonant electron assisted process, are
investigated (see FIG. 2). Particle 1 (1' and 1\textquotedblright ) is a
quasi-free electron of the metal, particle 2 is a quasi-free ingoing proton
and particle 4 is a localized proton. Particles 3 and 5 are different $Ni$
isotopes and they have mass numbers $A_{3}$ and $A_{5}$, respectively.
Particles 6 and 7 are $Cu$ isotopes of mass numbers $A_{3}+1$ and $A_{5}+1$,
respectively. Both nuclear transitions $2^{\prime },3\rightarrow 6$ and $%
4^{\prime },5\rightarrow 7$ are reactions of type $\left( \ref{NiAp}\right) $%
. The process is called quasi-resonant electron assisted double proton
capture process. Most of the daughter nuclei $_{29}^{A+1}Cu$ decay by the 
\begin{equation}
\text{ }_{29}^{A+1}Cu+e\rightarrow \text{ }_{28}^{A+1}Ni+Q_{EC}
\label{CuA1e}
\end{equation}%
electron capture reaction. TABLE I. of \cite{KaKe} contains the relevant
data for reactions $\left( \ref{NiAp}\right) $ and $\left( \ref{CuA1e}%
\right) $. As it is discussed above at a particular choice of the
participants resonances can be found in the processes of FIG. 2. The
resonance appears in line 4'.

Now particle 4 is a localized proton. We take $\psi _{4i}\left( \mathbf{x}%
\right) $ (see $\left( \ref{psziki}\right) $) as initial state of particle
4. The parameter $\beta _{4}$ in the case of a localized proton is $\beta
_{4}^{p}=\sqrt{m_{p}\omega _{4}/\hbar }$, where $m_{p}$ is the proton rest
mass and $\omega _{4}$ is the angular frequency of the ground state of a
3-dimensional harmonic oscillator of energy $E_{0}=\frac{3}{2}\hbar \omega
_{4}$. In $NiH_{0.75}$ the energy of an optical phonon $\hbar \omega _{4}=88$
$\left[ meV\right] $ \cite{Eckert}, that results $\beta _{4}^{p}=6.51\times
10^{8}$ $\left[ cm^{-1}\right] $ which is used in the calculation. The
Coulomb matrix elements which contain initial state of form $\psi
_{4i}\left( \mathbf{x}\right) $ also preserves momentum \cite{KaKe}. The
number density $N_{345}/V=4d^{-3}r_{A_{3}}r_{A_{5}}u$ (see $\left( \ref{N345}%
\right) $), where $u$ denotes the proton over metal number density and $d$
is the length of the $Ni$ elementary cell ($d=3.52\times 10^{-8}$ $\left[ cm%
\right] $).

Now we reformulate the results of section II. B. $\Delta _{A_{3}}=\Delta
_{02}+\Delta _{03}-\Delta _{06}$ is the energy of reaction $2^{\prime
}+3\rightarrow 6$ and $\Delta _{A_{5}}=\Delta _{04}+\Delta _{05}-\Delta
_{07} $ is the energy of reaction $4^{\prime }+5\rightarrow 7$. The $\Delta
_{0j}$-s are the energy defects of the corresponding nuclei and the total
reaction energy 
\begin{equation}
\Delta _{A_{3},A_{5}}=\Delta _{A_{3}}+\Delta _{A_{5}}.  \label{deltat5}
\end{equation}

It was found above that 
\begin{equation}
\delta _{A_{3},A_{5}}=\frac{\left( A_{3}+1\right) \Delta _{A_{3}}-\left(
A_{5}+1\right) \Delta _{A_{5}}}{A_{3}+A_{5}+2}  \label{deltab2a}
\end{equation}%
is a crucial quantity (it was defined by $\left( \ref{deltab}\right) $),
since, if $\delta _{A_{3},A_{5}}>0$ then resonance appears in line 4' at 
\begin{equation}
k_{4;A_{3},A_{5}}^{r}=\frac{\sqrt{2m_{4}\delta _{A_{3},A_{5}}}}{\hbar }.
\label{k4res2a}
\end{equation}

Substituting 
\begin{equation}
h_{corr,3}^{2}h_{corr,5}^{2}=\left( A_{3}-28\right) ^{2}(A_{5}-28)^{2}
\end{equation}%
into $\left( \ref{Wtot4}\right) $ the total rate of the 4th-order,
resonance-like electron assisted double proton capture processes has the form%
\begin{eqnarray}
W_{tot}^{(4)}\left( A_{3},A_{5}\right)  &=&\frac{4}{d^{3}}K_{0}^{\left(
4\right) }\left( A_{3}-28\right) ^{2}(A_{5}-28)^{2} \\
&&\times \left\langle F_{G}\right\rangle \chi N_{1}N_{2}r_{A_{3}}r_{A_{5}}u.
\notag  \label{Wtot42a}
\end{eqnarray}%
Here $N_{1}$ is the instantaneous number of the quasi-free electrons that
are felt by an ingoing particle 2, $N_{2}$ is the number of quasi-free
ingoing protons that can interact with particle triplets 3, 4 and 5, and $%
r_{A_{3}}$ and $r_{A_{5}}$ are the relative natural abundances of $Ni$
isotopes of mass numbers $A_{3}$ and $A_{5}$, respectively (see TABLE I. of 
\cite{KaKe}). $K_{0}^{(4)}$ given by $\left( \ref{K04}\right) $, where $%
R=1.2\times 10^{-13}\left[ cm\right] $ is the proton-radius. Taking $%
E_{1,Max}=0.1$ $\left[ MeV\right] $ again one gets 
\begin{equation}
4d^{-3}K_{0}^{\left( 4\right) }=1.9\times 10^{-7}\left[ MeV^{6}s^{-1}\right] 
\end{equation}%
in the case of $Ni$. Since $\mu _{23}\simeq \mu _{45}$ and $%
z_{2}z_{3}=z_{4}z_{5}=28$ therefore $F_{23}\left(
k_{4;A_{3},A_{5}}^{r}\right) \simeq F_{45}\left(
k_{4;A_{3},A_{5}}^{r}\right) $ in $\left\langle F_{G}\right\rangle $.
Furthermore, since particle 1 is an electron therefore $%
F_{12}(2k_{4;A_{1},A_{2}}^{r})=1$ and $\left( \ref{FG}\right) $ reads 
\begin{equation}
\left\langle F_{G;A_{3},A_{5}}\right\rangle =\left\langle
F_{12}(k_{i})\right\rangle F_{23}^{2}\left( k_{4;A_{3},A_{5}}^{r}\right) .
\label{FG2a}
\end{equation}%
Here $\left\langle F_{12}(k_{i}(E_{i}))\right\rangle =46.36\times \left(
E_{F}^{-1/2}\left[ eV\right] \right) $, where $E_{F}$ denotes the Fermi
energy. If $2m_{p}c^{2}\gg \delta $, true in our case, then $\left( \ref{khi}%
\right) $ can be written as 
\begin{equation}
\chi _{A_{3},A_{5}}=\frac{(A_{5}+1)^{3/2}\sqrt{2m_{0}c^{2}\Delta
_{A_{3},A_{5}}}}{4\delta _{A_{3},A_{5}}^{7}}.  \label{khi2a}
\end{equation}%
Here $m_{0}c^{2}=931.494$ $\left[ MeV\right] $ is the atomic mass unit.

One can see from TABLE I. of \cite{KaKe} that $\Delta _{A}$ increases
monotonically with the increase of $A$. Thus $\delta _{A_{3},A_{5}}>0$ if $%
A_{3}>A_{5}$, and the total power can be written as 
\begin{equation}
P_{tot}=N_{1}N_{2}uP_{0}\sum_{A_{3}>A_{5}}\psi _{A_{3},A_{5}},
\end{equation}%
where $P_{0}=4d^{-3}K_{0}^{\left( 4\right) }C_{E}\Delta _{0}^{-5}$. $%
C_{E}=1.602\times 10^{-13}$ $\left[ J/MeV\right] $ is the energy unit
conversion factor, $\Delta _{0}=1$ $\left[ MeV\right] $ as an order of
magnitude of a typical nuclear reaction energy value. The use of $\Delta _{0}
$ makes%
\begin{eqnarray}
\psi _{A_{3},A_{5}} &=&\left( A_{3}-28\right)
^{2}(A_{5}-28)^{2}r_{A_{3}}r_{A_{5}}\times  \\
&&\times \Delta _{A_{3},A_{5}}\left\langle F_{G;A_{3},A_{5}}\right\rangle
\chi _{A_{3},A_{5}}\Delta _{0}^{5}  \notag
\end{eqnarray}%
dimensionless. In our case 
\begin{equation}
P_{0}=3.1\times 10^{-20}\left[ W\right] 
\end{equation}%
and%
\begin{equation}
\sum_{A_{3}>A_{5}}\psi _{A_{3},A_{5}}=6.48\times 10^{-6}\times \left(
E_{F}^{-1/2}\left[ eV\right] \right) .
\end{equation}%
In $\sum_{A_{3}>A_{5}}\psi _{A_{3},A_{5}}$ the significant contributions are
given by $\psi _{61,58}$, $\psi _{62,58}$, $\psi _{64,58}$ and $\psi _{64,60}
$, with $\psi _{64,58}=6.36\times 10^{-6}\times \left( E_{F}^{-1/2}\left[ eV%
\right] \right) $ as the leading term responsible for 98\% of the effect. A
hydrogen concentration independent $E_{F}=17\left[ eV\right] $ is used
producing 
\begin{equation}
P_{tot}=4.8\times 10^{-26}\times N_{1}N_{2}u\left[ W\right] .  \label{Ptot2}
\end{equation}

Now we proceed to the determination of $N_{1}$, $N_{2}$ and $u$. $N_{1}$
stands for the number of valence electrons which can interact with a proton
(particle 2). Considering that $Ni$ consists of micro crystals of linear
dimension of about $\left( D\left[ \mu m\right] \right) $ $1-10$ $\mu m$, it
is reasonable to assume that a proton penetrating the material "feels" all
the valence electrons. Thus%
\begin{equation}
N_{1}=\frac{V_{g}}{v_{c}}g_{e},
\end{equation}%
where $V_{g}=D^{3}\times 10^{-12}\left[ cm^{3}\right] $ is the volume of
micro crystals, $v_{c}=d^{3}/4$ is the volume of the elementary cell and $%
g_{e}=10$ is the number of valence electrons in an elementary cell of $Ni$.
From it $N_{1}=D^{3}\left[ \mu m\right] \times 9.17\times 10^{11}$. $N_{2}$
denotes the number of protons which can be considered quasi free in respect
of the process. It can be calculated if the number density $n_{p}$ in gas is
multiplied with the metal volume $V_{2}$, where they are taken for free. The
volume is the product of the surface $F$ of the sample and the length $d$ of
the elementary cell. The result is 
\begin{equation}
N_{2}=2Fn_{L}\frac{pT_{0}}{p_{0}T}d,  \label{N2}
\end{equation}%
where $n_{L}=2.69\times 10^{19}\left[ cm^{-3}\right] $ is the number density
in the gas of pressure $p_{0}=1\left[ atm\right] $ and temperature $T_{0}=273%
\left[ ^{0}K\right] $. The actual pressure and temperature are denoted by $p$
and $T$. Factor 2 follows from the fact that the hydrogen molecule contains
2 atoms. Here the catalytic process producing atomic hydrogen is not
considered, it supposed, which is a rough over estimation, that at the $Ni$
surface the whole gas is atomistic.

In \cite{Focardi1} excess heat power $44\left[ W\right] $ is obtained from a 
$Ni$ rod of diameter $d_{0}=0.5\left[ cm\right] $ and length $h_{0}=9\left[
cm\right] $ at $p=0.5\left[ atm\right] $ and $T=753\left[ ^{0}K\right] $. At
this temperature and pressure $u\simeq 7.9\times 10^{-5}$ \cite{Fast}.
Utilizing this and the values $N_{1}$ and $N_{2}$ obtained above, $%
P_{tot}=D^{3}\times 1.2\times 10^{-6}\left[ W\right] $ from $\left( \ref%
{Ptot2}\right) $. In case of $D\simeq 330\left[ \mu m\right] $ the result is 
$44\left[ W\right] $, which considering that in this simple model a great
number of solid state processes were neglected and in the nuclear processes
it was only the Weisskopf approximation in which the matrix element was
calculated, is a very good approximation.

\section{Summary}

Resonance-like heavy particle and electron assisted double nuclear processes
in solids are discussed. The transition probabilities per unit time of the
3rd-order heavy particle assisted and the 4th-order electron assisted
resonance-like double nuclear processes are determined. The 3rd-order heavy
particle assisted and the 4th-order electron assisted resonance-like double
nuclear processes may partly be responsible for the so called anomalous
screening effect observed in low energy accelerator physics investigating
astrophysical factors of nuclear reactions of low atomic numbers \cite%
{Raiola1}. The theoretical description of the doubled processes discussed
extends the possible explanation and description of LENR with its nuclear
physical background. It is found that the $d+d\rightarrow $ $^{4}He$ process
coupled to the $_{46}^{A}Pd+d\rightarrow $ $_{47}^{A+1}Ag$ process due to
the quasi-resonant electron assisted doubled nuclear process has extremely
large rate. The $_{2}^{4}He$ production with $34.7$ $MeV/He$ obtained in the
leading, 4th-order, quasi-resonant electron assisted $d+d\rightarrow $ $%
_{2}^{4}He$ process fits well with the observed $32\pm 13$ $MeV/He$ value of
LENRs \cite{Storms2}.

With the help of our theory describing resonance-like electron assisted
doubled nuclear processes we estimated the nuclear power $\left( \ref{Ptot4}%
\right) $ created in the $d+d\rightarrow $ $^{4}He$ process coupled to the $%
_{46}^{A}Pd+d\rightarrow $ $_{47}^{A+1}Ag$ process due to the quasi-resonant
electron assisted doubled nuclear process and the nuclear power $\left( \ref%
{Ptot2}\right) $ due to double proton capture in $Ni-H$ systems formed by $%
Ni $ placed in $H_{2}$ gas environment. The nuclear powers are consistent
with observations. Moreover, the magnitude of the power obtained in $Ni-H$
systems containing $Ni$\ in powdered form is considerable from practical
point of view.

The authors are indebted to K. H\"{a}rtlein for his technical assistance.

\section{Appendix I. Rate of resonance-like heavy charged particle assisted
doubled nuclear processes}

The rate of the process is 
\begin{equation}
W_{fi}^{(3)}=\frac{2\pi }{\hbar }\sum_{f}\left\vert T_{if}^{(3)}\right\vert
^{2}\delta (E_{f}-\Delta )  \label{Wfi3}
\end{equation}%
where%
\begin{align}
T_{if}^{(3)}& =\sum_{\nu }\sum_{\mu }\frac{V_{f\nu }^{St}V_{\nu \mu
}^{St}V_{\mu i}^{Cb}}{\left( E_{\nu }-E_{i}\right) \left( E_{\mu
}-E_{i}\right) }  \label{Tfi3} \\
& \times \frac{\left( 2\pi \right) ^{3}}{V}\delta \left( \mathbf{k}_{5}+%
\mathbf{k}_{6}\right) .  \notag
\end{align}%
$\mathbf{k}_{j}$ denotes the wave vector of particle $j$ in the final state.
It will be seen that the process may have resonance like character if the
masses of particles 5 and 6 differ significantly, therefore the $T_{if}^{(3)}
$ attached to the leading graph (e.g. FIG. 1(a) in the case $m_{5}\gg m_{6}$
discussed) is enough to calculate. $\mathbf{k}_{jn}$ denotes the wave vector
of particle $j$ in state $n=i,\mu $ or $\nu $. The initial wave vector $%
\mathbf{k}_{2i}$ of particle 2 is neglected in the Coulomb matrix element 
\begin{equation}
V_{\mu i}^{Cb}=g_{S}\frac{4\pi e^{2}z_{1}z_{2}}{V^{3/2}}\frac{\widetilde{%
\psi }_{1i}\left( \mathbf{k}_{1\mu }+\mathbf{k}_{2\mu }\right) }{\left\vert 
\mathbf{k}_{2\mu }\right\vert ^{2}+\lambda ^{2}},  \label{VCbmui}
\end{equation}%
where the 
\begin{equation}
\widetilde{\psi }_{1i}(\mathbf{k}_{1\mu }+\mathbf{k}_{2\mu })=8\pi
^{9/4}\beta _{1}^{3/2}\delta (\mathbf{k}_{1\mu }+\mathbf{k}_{2\mu })
\label{pszikal}
\end{equation}%
formula can be used assuming that particle 1 is localized. $\widetilde{\psi }%
_{1i}$ stands for the Fourier transform of the initial state $\psi _{1i}$.

The two nuclear matrix elements 
\begin{align}
V_{\nu \mu }^{St,W}\left( a\right) & =-2f^{2}f_{14}(k_{1\mu })\sqrt{\frac{%
12\pi R}{V}}\left( 1-\frac{2}{e}\right) \times  \label{VStnumu} \\
& \times \frac{\left( 2\pi \right) ^{3}}{V}\delta \left( \mathbf{k}_{1\mu }-%
\mathbf{k}_{1\nu }\right) ,  \notag
\end{align}%
and 
\begin{equation}
V_{f\nu }^{St,W}\left( a\right) =-2f^{2}f_{23}(k_{2\nu })\sqrt{\frac{12\pi R%
}{V}}\left( 1-\frac{2}{e}\right)  \label{VStfnu}
\end{equation}%
are valid in the case of the process of FIG. 1(a) and 
\begin{align}
V_{\nu \mu }^{St,W}\left( b\right) & =-2f^{2}f_{23}(k_{2\mu })\sqrt{\frac{%
12\pi R}{V}}\left( 1-\frac{2}{e}\right)  \label{VStnumub} \\
& \times \frac{\left( 2\pi \right) ^{3}}{V}\delta \left( \mathbf{k}_{1\mu }-%
\mathbf{k}_{1\nu }\right) ,  \notag
\end{align}%
\begin{equation}
V_{f\nu }^{St,W}\left( b\right) =-2f^{2}f_{14}(k_{1\nu })\sqrt{\frac{12\pi R%
}{V}}\left( 1-\frac{2}{e}\right)  \label{VStfnub}
\end{equation}%
stand for the process of FIG. 1(b). Here $R$ is the radius of a nucleon (we
take $R=1.2\times 10^{-13}\left[ cm\right] $, that is the proton-radius) and
the single nucleon approach in the Weisskopf approximation is used. The
energy differences in the denominator of $\left( \ref{Tfi3}\right) $ are:
the difference of the kinetic energies%
\begin{equation}
E_{\mu }-E_{i}=\frac{\hbar ^{2}k_{1\mu }^{2}}{2m_{1}}+\frac{\hbar
^{2}k_{2\mu }^{2}}{2m_{2}}-E_{i,kin}  \label{Enumu}
\end{equation}%
and 
\begin{equation}
\left( E_{\nu }-E_{i}\right) _{a}=\frac{\hbar ^{2}k_{2\nu }^{2}}{2m_{2}}+%
\frac{\hbar ^{2}k_{5}^{2}}{2m_{5}}-E_{i,kin}-\Delta _{5}  \label{Enuia}
\end{equation}%
in the case of FIG. 1(a) and 
\begin{equation}
\left( E_{\nu }-E_{i}\right) _{b}=\frac{\hbar ^{2}k_{1\nu }^{2}}{2m_{1}}+%
\frac{\hbar ^{2}k_{6}^{2}}{2m_{6}}-E_{i,kin}-\Delta _{6}  \label{Enuib}
\end{equation}%
in the case of FIG. 1(b). In $\left( \ref{Enuia}\right) $ and $\left( \ref%
{Enuib}\right) $ the rest energy differences have to be also taken into
account because of the nuclear reaction $1^{\prime }+4$ $\rightarrow 5$ in
the case of (a) and $2^{\prime }+3\rightarrow 6$ in the case of (b). $\Delta
_{5}=\Delta _{01}+\Delta _{04}-\Delta _{05}$ is the energy of reaction $%
1^{\prime }+4$ $\rightarrow 5$, and $\Delta _{6}=\Delta _{02}+\Delta
_{03}-\Delta _{06}$ is the energy of reaction $2^{\prime }+3\rightarrow 6$.
The $\Delta _{0j}$-s are the energy defects of the corresponding nuclei. The
total reaction energy 
\begin{equation}
\Delta =\Delta _{5}+\Delta _{6}.  \label{Delta}
\end{equation}%
After performing the $\sum_{f}$ (really using the $\sum_{f}$ $\rightarrow $ $%
\int \left[ V/\left( 2\pi \right) ^{3}\right] d\mathbf{k}_{5}\times \int %
\left[ V/\left( 2\pi \right) ^{3}\right] d\mathbf{k}_{6}$ correspondence and
integrating over $\mathbf{k}_{6}$) in $\left( \ref{Wfi3}\right) $ the Dirac
delta $\delta \left( \mathbf{k}_{5}+\mathbf{k}_{6}\right) $ in $\left( \ref%
{Tfi3}\right) $ will result $\mathbf{k}_{6}=-\mathbf{k}_{5}$, and the final
energy can be written as $E_{f}=$ $\hbar ^{2}k_{5}^{2}/\left( 2\mu
_{56}\right) $. The energy Dirac delta will result 
\begin{equation}
k_{5}^{2}=\frac{2\mu _{56}\Delta }{\hbar ^{2}}.  \label{k52}
\end{equation}%
Furthermore because of the presence of $\delta \left( \mathbf{k}_{2\mu }-%
\mathbf{k}_{2\nu }\right) $ and $\delta \left( \mathbf{k}_{1\mu }-\mathbf{k}%
_{1\nu }\right) $ in the matrix elements $V_{\nu \mu }^{St,W}\left( a\right) 
$ and $V_{\nu \mu }^{St,W}\left( b\right) $ the $\sum_{\nu }$ $\rightarrow $ 
$\int \left[ V/\left( 2\pi \right) ^{3}\right] d\mathbf{k}_{2\nu }$ and $%
\sum_{\nu }$ $\rightarrow $ $\int \left[ V/\left( 2\pi \right) ^{3}\right] d%
\mathbf{k}_{1\nu }$ will allow the $\mathbf{k}_{2\nu }=\mathbf{k}_{2\mu }$
and $\mathbf{k}_{1\nu }=\mathbf{k}_{1\mu }$ substitutions in them and in the
energy denominators as well. The initial kinetic energy $E_{i,kin}$ is
neglected in $\left( E_{\nu }-E_{i}\right) _{a}$ and $\left( E_{\nu
}-E_{i}\right) _{b}$. Thus%
\begin{equation}
\left( E_{\nu }-E_{i}\right) _{a}=\frac{\hbar ^{2}k_{2\mu }^{2}}{2m_{2}}%
+\delta _{a}  \label{Enuia2}
\end{equation}%
with%
\begin{equation}
\delta _{a}=\frac{\mu _{56}}{m_{5}}\Delta -\Delta _{5}  \label{da}
\end{equation}%
and%
\begin{equation}
\left( E_{\nu }-E_{i}\right) _{b}=\frac{\hbar ^{2}k_{1\mu }^{2}}{2m_{1}}%
+\delta _{b}  \label{Enuib2}
\end{equation}%
with%
\begin{equation}
\delta _{b}=\frac{\mu _{56}}{m_{6}}\Delta -\Delta _{6}.  \label{db}
\end{equation}%
Since%
\begin{equation}
\delta _{a}+\delta _{b}=0  \label{dadb0}
\end{equation}%
one of $\delta _{a}$ and $\delta _{b}$ is negative. Let us suppose that $%
\delta _{a}<0$. It means if 
\begin{equation}
\frac{\hbar ^{2}k_{2\mu }^{2}}{2m_{2}}=-\delta _{a}=\Delta _{5}-\frac{\mu
_{56}}{m_{5}}\Delta  \label{res}
\end{equation}%
then $\left( E_{\nu }-E_{i}\right) _{a}=0$ and we find that the process (a)
has resonance-like behavior when $k_{2\mu }=k^{r}$ with $k^{r}=\sqrt{%
2m_{2}\left( \Delta _{5}-\frac{\mu _{56}}{m_{5}}\Delta \right) }/\hbar $. If 
$\delta _{a}<0$ then $\delta _{b}>0$ and the process (b) can not have
resonance character, and therefore it is enough to calculate the $%
T_{if}^{(3)}$ attached to graph 3(a).

Let us introduce the half width $\Gamma $ of the resonance with which the
complex energy difference, denoted by suffix $C$, reads 
\begin{equation}
\left( E_{\nu }-E_{i}\right) _{a,C}=\frac{\hbar ^{2}k_{2\mu }^{2}}{2m_{2}}%
+\delta _{a}-i\frac{\Gamma }{2}  \label{Enuia3}
\end{equation}%
that equals $-i\frac{\Gamma }{2}$ if the resonance condition $\left( \ref%
{res}\right) $ is met. With the use of the correspondence $\sum_{\mu }$ $%
\rightarrow $ $\int \left[ V/\left( 2\pi \right) ^{3}\right] d\mathbf{k}%
_{2\mu }$ $\times $ $\int \left[ V/\left( 2\pi \right) ^{3}\right] d\mathbf{k%
}_{1\mu }$ in $\left( \ref{Tfi3}\right) $ and after carrying out the
integral over $\mathbf{k}_{1\mu }$ the relation $\left( \ref{pszikal}\right) 
$ gives $\mathbf{k}_{1\mu }=-\mathbf{k}_{2\mu }$. Integrating over $\mathbf{k%
}_{2\mu }$ we have the integral of the form%
\begin{equation}
I=\int \frac{h\left( k_{2\mu }\right) d\mathbf{k}_{2\mu }}{\frac{\hbar
^{2}k_{2\mu }^{2}}{2m_{2}}+\delta _{a}-i\frac{\Gamma }{2}},  \label{I}
\end{equation}%
where $h\left( k_{2\mu }\right) $ is any function of $k_{2\mu }$, that in
this case 
\begin{equation}
h\left( k_{2\mu }\right) =g_{S}\left( k_{2i},2k_{2\mu }\right)
f_{23}(k_{2\mu })f_{14}(k_{2\mu }).  \label{hk2mu}
\end{equation}%
For evaluating $\left( \ref{I}\right) $ the $g\left( k_{2\mu }\right) =\frac{%
\hbar ^{2}k_{2\mu }^{2}}{2m_{2}}+\delta _{a}$ is introduced and the
substitution 
\begin{equation}
\frac{1}{g\left( k_{2\mu }\right) -i\varepsilon }\rightarrow P\frac{1}{%
g\left( k_{2\mu }\right) }+i\pi \delta \left( g\left( k_{2\mu }\right)
\right)  \label{xiep}
\end{equation}%
is applied resulting 
\begin{equation}
I=I_{P}+iI_{\delta },  \label{I2}
\end{equation}%
where 
\begin{equation}
I_{P}=\int P\frac{h\left( k_{2\mu }\right) d\mathbf{k}_{2\mu }}{\frac{\hbar
^{2}k_{2\mu }^{2}}{2m_{2}}+\delta _{a}}  \label{IP}
\end{equation}%
and 
\begin{equation}
I_{\delta }=\pi \int h\left( k_{2\mu }\right) \delta \left( g\left( k_{2\mu
}\right) \right) d\mathbf{k}_{2\mu }.  \label{Idelta}
\end{equation}%
Using the identity 
\begin{equation}
\delta \left( g\left( k_{2\mu }\right) \right) =\frac{\delta \left( k_{2\mu
}-k^{r}\right) }{\frac{d}{dk_{2\mu }}g\left( k_{2\mu }\right) \left.
{}\right\vert _{k_{2\mu }=k^{r}}}  \label{dgk2mu}
\end{equation}%
where $k^{r}$ (see $\left( \ref{kres}\right) $) is the root of the $g\left(
k_{2\mu }\right) =0$ equation (see $\left( \ref{res}\right) $). In a lower
estimation of $I$ as $\left\vert I\right\vert =I_{\delta }$, the relevant
part of $\left( \ref{Tfi3}\right) $ is approximated as 
\begin{align}
& \left\vert \int \frac{h\left( k_{2\mu }\right) }{\frac{\hbar ^{2}k_{2\mu
}^{2}}{2m_{2}}+\delta _{a}-i\frac{\Gamma }{2}}d\mathbf{k}_{2\mu }\right\vert
\label{resint} \\
& =4\pi ^{2}k^{r}m_{2}h\left( k^{r}\right) /\hbar ^{2}.  \notag
\end{align}%
Applying the above relations and results, a lower approximation of the
transition rate reads as 
\begin{equation}
W_{fi}^{(3)}=K^{(3)}K_{1}^{(3)}\left( \Delta ,\Delta _{5}\right) H\left(
k^{r}\right) h_{corr,3}^{2}h_{corr,4}^{2}  \label{Wfi30}
\end{equation}%
where $K^{(3)}$, $K_{1}^{(3)}\left( \Delta ,\Delta _{5}\right) $ and $%
H\left( k^{r}\right) $ are given by $\left( \ref{K3}\right) $, $\left( \ref%
{K13}\right) $ and $\left( \ref{Hkr}\right) $, and for $h_{corr,3}$ and $%
h_{corr,5}$ see $\left( \ref{hcorrk}\right) $ and $\left( \ref{hcorrk2}%
\right) $, that are defined by Eq.(45) of \cite{KaKe}, and $k^{r}$ is
determined by $\left( \ref{kres}\right) $.

\section{Appendix II. Rate of resonance-like electron assisted doubled
nuclear processes}

The rate of the process 
\begin{equation}
W_{fi}^{(4)}=\frac{2\pi }{\hbar }\sum_{f}\left\vert T_{if}^{(4)}\right\vert
^{2}\delta (E_{f}-\Delta )  \label{Wfi4}
\end{equation}%
with%
\begin{equation}
T_{if}^{(4)}=T_{if}^{(4)}(a)+T_{if}^{(4)}(b)+T_{if}^{(4)}(c),  \label{Tfi4}
\end{equation}%
where $T_{if}^{(4)}(a)$, $T_{if}^{(4)}(b)$ and $T_{if}^{(4)}(c)$ are the
matrix elements of the processes of FIG. 2 (a), (b) and (c), respectively.
If processes of FIG. 2 (a) and (b) have resonance-like character (see later)
then the process of FIG. 2 (c) has not and therefore its contribution may be
neglected. Thus we take into account the contributions of 
\begin{align}
T_{if}^{(4)}\left( a\right) & =\sum_{\rho }\sum_{\nu }\sum_{\mu }\frac{%
V_{f\rho }^{St}V_{\rho \nu }^{Cb}V_{\nu \mu }^{St}}{\left( E_{\rho
}-E_{i}\right) \left( E_{\nu }-E_{i}\right) }  \label{Tfi4a} \\
& \times \frac{V_{\mu i}^{Cb}}{\left( E_{\mu }-E_{i}\right) }\frac{\left(
2\pi \right) ^{3}}{V}\delta \left( \mathbf{k}_{1f}+\mathbf{k}_{6}+\mathbf{k}%
_{7}\right)  \notag
\end{align}%
and 
\begin{align}
T_{if}^{(4)}\left( b\right) & =\sum_{\rho }\sum_{\nu }\sum_{\mu }\frac{%
V_{f\rho }^{St}V_{\rho \nu }^{St}V_{\nu \mu }^{Cb}}{\left( E_{\rho
}-E_{i}\right) \left( E_{\nu }-E_{i}\right) }  \label{Tfi4b} \\
& \times \frac{V_{\mu i}^{Cb}}{\left( E_{\mu }-E_{i}\right) }\frac{\left(
2\pi \right) ^{3}}{V}\delta \left( \mathbf{k}_{1f}+\mathbf{k}_{6}+\mathbf{k}%
_{7}\right) .  \notag
\end{align}%
The outline of the calculation is the following. For the Coulomb matrix
element of the process $1,2$ $\rightarrow 1^{\prime },2^{\prime }$ we use
the form $V_{\mu i}^{Cb}$ given by Eq.(37) of \cite{KaKe}. For calculating
the matrix element of the Coulomb interaction of the process $1^{\prime
},4\rightarrow 1^{\prime \prime },4^{\prime }$ the form given by $\left( \ref%
{VCbmui}\right) $ and the approximation $\left( \ref{pszikal}\right) $ are
used. Thus in each filled dot of the graphs representing a Coulomb
interaction the momentum (wave number) is conserved. The initial wave
vectors of particles 1, 2 and 4 are neglected. The matrix elements of
nuclear transitions $2^{\prime },3\rightarrow 6$ and $4^{\prime
},5\rightarrow 7$ are calculated in the Weisskopf approximation applying the
appropriate one from formulae $\left( \ref{VStnumu}\right) $, $\left( \ref%
{VStnumub}\right) $, $\left( \ref{VStfnu}\right) $ and $\left( \ref{VStfnub}%
\right) $ with the appropriate $f_{23}\left( k_{2}\right) $ and $%
f_{45}\left( k_{4}\right) $ functions in it, respectively. The initial
motion of particles 3 and 5, i.e. their initial wave vectors are also
neglected.\ In summing up for the intermediate and final states and for the
square of the Dirac delta of argument of wave vector the correspondences and
relations used above are applied again. (Remember, that now $\Delta
_{6}=\Delta _{02}+\Delta _{03}-\Delta _{06}$ is the energy of reaction $%
2^{\prime }+3\rightarrow 6$ and $\Delta _{7}=\Delta _{05}+\Delta
_{04}-\Delta _{07}$ is the energy of reaction $4^{\prime }+5$ $\rightarrow 7$%
. The $\Delta _{0j}$-s are again the energy defects of the corresponding
nuclei and the total reaction energy $\Delta =\Delta _{6}+\Delta _{7}$). We
calculate the rate of those processes in which the kinetic energy $E_{1f}$
of the electron can be neglected in the energy $E_{f}$ of the final state.
Consequently the $V/\left( 2\pi \right) ^{3}\int d\mathbf{k}_{1f}$ will
result a factor $V/\left( 2\pi \right) ^{3}\left( 4\pi /3\right)
k_{1,Max}^{3}$, where $k_{1,Max}$ is the maximum of the possible wave
vectors of the electron in the final state. Furthermore, in the energy Dirac
delta 
\begin{equation}
E_{f}=E_{6}+E_{7}-\Delta  \label{Ef4}
\end{equation}%
is used. Neglecting also the final wave vector $\mathbf{k}_{1f}$ of the
electron in the Dirac delta, $\delta (\mathbf{k}_{6}+\mathbf{k}_{7})$ is
used in $\left( \ref{Tfi4a}\right) $, $\left( \ref{Tfi4b}\right) $ resulting%
\begin{equation}
E_{f}=\frac{\hbar ^{2}k_{7}^{2}}{2\mu _{67}}-\Delta  \label{Ef42}
\end{equation}%
after integration over $\mathbf{k}_{6}$ ($k_{6}=k_{7}$).

Since in the cases investigated $\mathbf{k}_{1f}$ is neglected, the wave
number vector conservation in Coulomb scattering results the conservation of
the magnitude of wave vector in lines 1', 2' and 4', and it is denoted by $%
k_{4}$. Let us now investigate the energy denominators. The intermediate
states are labeled with $\mu $, $\nu $ and $\rho $. In the case of graph (a)%
\begin{equation}
\left( E_{\mu }-E_{i}\right) _{a}=\frac{\hbar ^{2}k_{4}^{2}}{2m_{2}}+\sqrt{%
\left( \hbar ck_{4}\right) ^{2}+m_{e}^{2}c^{4}}-m_{e}c^{2},  \label{Emuia4}
\end{equation}%
\begin{equation}
\left( E_{\nu }-E_{i}\right) _{a}=\frac{\hbar ^{2}k_{7}^{2}}{2m_{6}}-\Delta
_{6}+\sqrt{\left( \hbar ck_{4}\right) ^{2}+m_{e}^{2}c^{4}}-m_{e}c^{2},
\label{Enuia4}
\end{equation}%
and 
\begin{equation}
\left( E_{\rho }-E_{i}\right) _{a}=\frac{\hbar ^{2}k_{7}^{2}}{2m_{6}}-\Delta
_{6}+\frac{\hbar ^{2}k_{4}^{2}}{2m_{4}}.  \label{Eroia4}
\end{equation}%
In obtaining $\left( \ref{Eroia4}\right) $ $E_{1f}$ is neglected. In the
case of graph (b) 
\begin{equation}
\left( E_{\mu }-E_{i}\right) _{b}=\left( E_{\mu }-E_{i}\right) _{a},
\label{Emuib4}
\end{equation}%
\begin{equation}
\left( E_{\nu }-E_{i}\right) _{b}=\frac{\hbar ^{2}k_{4}^{2}}{2m_{2}}+\frac{%
\hbar ^{2}k_{4}^{2}}{2m_{4}},  \label{Enuib4}
\end{equation}%
and 
\begin{equation}
\left( E_{\rho }-E_{i}\right) _{b}=\frac{\hbar ^{2}k_{7}^{2}}{2m_{6}}-\Delta
_{6}+\frac{\hbar ^{2}k_{4}^{2}}{2m_{4}}.  \label{Eroib4}
\end{equation}%
One can see from $\left( \ref{Eroia4}\right) $ and $\left( \ref{Eroib4}%
\right) $ that $\left( E_{\rho }-E_{i}\right) _{a}=\left( E_{\rho
}-E_{i}\right) _{b}$. Integrating over $\mathbf{k}_{7}$ and using the energy
Dirac delta%
\begin{equation}
\left( E_{\rho }-E_{i}\right) _{a}=\left( E_{\rho }-E_{i}\right) _{b}=\frac{%
\hbar ^{2}k_{4}^{2}}{2m_{4}}-\delta _{a,b}  \label{res4}
\end{equation}%
with%
\begin{equation}
\delta _{a,b}=\Delta _{6}-\frac{\mu _{67}}{m_{6}}\Delta =\Delta _{6}-\frac{%
m_{7}}{m_{6}+m_{7}}\Delta .  \label{deltab}
\end{equation}%
If $\delta _{a,b}>0$ then in $\left( \ref{Tfi4a}\right) $ and $\left( \ref%
{Tfi4b}\right) $ resonance appears at $k_{4}^{r}$ given by $\left( \ref%
{k4res}\right) $. Let us introduce again the half width $\Gamma $ of the
resonance with which the complex energy differences read 
\begin{equation}
\left( E_{\rho }-E_{i}\right) _{a,C}=\left( E_{\rho }-E_{i}\right) _{b,C}=%
\frac{\hbar ^{2}k_{4}^{2}}{2m_{4}}-\delta _{a,b}-i\frac{\Gamma }{2}.
\label{res4g6}
\end{equation}%
The integration over $\mathbf{k}_{4}$ 
\begin{align}
& \left\vert \int \frac{f_{23}(k_{4})f_{45}(k_{4})}{\frac{\hbar ^{2}k_{4}^{2}%
}{2m_{4}}-\delta _{a,b}-i\frac{\Gamma }{2}}d\mathbf{k}_{4}\right\vert
\label{resint4} \\
& =4\pi ^{2}k_{4}^{r}m_{4}f_{23}(k_{4}^{r})f_{14}(k_{4}^{r})/\hbar ^{2}, 
\notag
\end{align}%
with the aid of $\left( \ref{resint}\right) $, where $k_{4}^{r}$ is given by 
$\left( \ref{k4res}\right) $. (Now, because particle 1 is an electron $%
g_{S}(k_{11^{\prime }},k_{4^{\prime }1^{\prime }})\simeq g_{S}(k_{1^{\prime
}},k_{4^{\prime }})=1$ .)

The energy differences in the denominators of $\left( \ref{Tfi4a}\right) $
and $\left( \ref{Tfi4b}\right) $ will be%
\begin{equation}
\left( E_{\nu }-E_{i}\right) _{a}=\delta _{a,b}\left( \sqrt{\frac{2m_{4}c^{2}%
}{\delta _{a,b}}}-1\right) =\Delta _{a},  \label{Enia4d}
\end{equation}%
\begin{equation}
\left( E_{\nu }-E_{i}\right) _{b}=\frac{m_{4}}{\mu _{24}}\delta
_{a,b}=\Delta _{b},  \label{Enuib4d}
\end{equation}%
and 
\begin{equation}
\left( E_{\mu }-E_{i}\right) _{b}=\left( E_{\mu }-E_{i}\right) _{a}=\delta
_{a,b}\left( \frac{m_{4}}{m_{2}}+\sqrt{\frac{2m_{4}c^{2}}{\delta _{a,b}}}%
\right) =\Delta _{ab}.  \label{deltab2}
\end{equation}%
Now the rate of the 4th-order processes is 
\begin{equation}
W_{fi}^{(4)}=K_{0}^{(4)}F_{G}\frac{\chi \left( \Delta \right) }{V}%
h_{corr,3}^{2}h_{corr,5}^{2}  \label{Wfi42}
\end{equation}%
where $K_{0}^{(4)}$, $F_{G}$ and $\chi \left( \Delta \right) $ are
determined by $\left( \ref{K04}\right) $, $\left( \ref{FG}\right) $ and $%
\left( \ref{khi}\right) $, $k_{4}^{r}$ is given by $\left( \ref{k4res}%
\right) $ and for $h_{corr,3}$ and $h_{corr,5}$ see $\left( \ref{hcorrk}%
\right) $ and $\left( \ref{hcorrk2}\right) $.

\end{document}